\documentclass[12pt, eqsecnums]{iopart}
\usepackage{graphicx, epsfig, iopams}

\newcommand{\be}{\begin{equation}}
\newcommand{\ee}{\end{equation}}
\newcommand{\bea}{\begin{eqnarray}}
\newcommand{\eea}{\end{eqnarray}}
\begin{document}

\title[Why did the Universe Start from a Low Entropy State?]{Why did the Universe Start from a Low Entropy State?
 Implications for Holography, Complementarity and Eternal Inflation}

\author{R.~Holman}
\address{Department of Physics, Carnegie Mellon University, Pittsburgh PA 15213, USA}
\ead{rh4a@andrew.cmu.edu}
\author{ L.~Mersini-Houghton}
\address{Department of Physics and Astrononmy, UNC-Chapel Hill, NC, 27599-3255, USA}
\ead{mersini@physics.unc.edu}

\date{\today}
 
\begin{abstract} 
In previous work, we showed that the answer to the question posed in the title cannot be found within an equilibrium setting. The inclusion of {\em dynamical} backreaction effects from massive long wavelength modes on the initial DeSitter patches results in gravitational instabilities that cleanse the phase space of inflationary initial conditions of all regions except those allowing high energy inflation. This interplay between the matter and gravitational degrees of freedom explains why ergodicity is broken and why the Universe starts in an out-of equlibrium state with low entropy. Here we argue that this reduction of the phase space of inflationary initial conditions implies that an inflationary universe is incompatible with holography inspired proposals such as causal patch physics and D (N)-bounds. We also discuss why chaotic or eternal inflation may not resolve the puzzle of the initial conditions and of the arrow of time.
\end{abstract}


\maketitle

\section{Introduction}
One of the main issues besetting the inflationary universe paradigm is that of initial conditions. How likely is it that, even given a slow roll potential that allows for inflation, that the inflaton will find itself in the part of field space that allows for inflation? This question becomes even more pointed given the WMAP data\cite{wmap}. It is consistent with the predictions from slow-roll inflation; furthermore these data seem to prefer inflation happening at a high energy scale, $H_{\rm inf}\sim 10^{14}-10^{16}$ GeV. We can estimate the probability for high scale inflation by computing the entropy $S_{\rm deS}$ for a de Sitter space with cosmological constant $\Lambda$: $S_{\rm deS} = 3\pi \slash G \Lambda$  \cite{gh}. The probability is then given by $P\propto \exp S_{\rm deS}$, which prefers small $\Lambda$ (or large horizon size $\Lambda^{-1}$), thus precluding high scale inflation. However, this probability is at odds with the existence of an arrow of time which requires a low entropy for the initial state. The question of whether the Universe was able to inflate can then be rephrased as: why did the Universe start from a low entropy state?

These results\cite{eqdesitter} have their roots in the assumption of an {\em equilibrium} ensemble of initial inflationary patches. This assumption is, in our view, not warranted. Gravitational systems have negative heat capacity which makes attaining statistical equilibrium difficult if not impossible. A more reasonable approach to the question of inflationary initial conditions would be a {\em dynamical} one. A brief account of such a dynamical mechanism is given in the next section. Details can be found in \cite{us1}. 

The purpose of the present letter is to investigate the implications of our quantum dynamical treatment for holography, the principle of complementarity and eternal inflation. This analysis is done by probing into the validity of the assumptions of equilibrium, ergodicity and causality, upon which these proposals are based. 

The inclusion of the backreaction from the quantum fluctuations due to both metric and scalar perturbations gives rise to instabilities that render most of the inflationary patches {\em unstable} against gravitational collapse of super-horizon modes. This has the effect of dynamically reducing the allowed phase space of stable inflationary patches. This is essentially a Jeans instability effect, arising from the generation of tachyonic modes by the backreaction of the perturbations in Wheeler-deWitt Master equation. We can then trace out the modes corresponding to collapsing patches to construct a reduced density matrix $\rho_{\rm red}$ for the patches that survive and enter an inflationary phase.  We use this to show explicitly that $\left[\hat{H},\rho_{\rm red}\right]\neq 0$, which clearly indicates that the initial states allowing for inflation do {\em not} form an equilibrium ensemble. 

Our analysis is tied into current efforts\cite{selection} to select appropriate vacua from the landscape of string vacua\cite{landscape}. The landscape minisuperspace becomes the phase space of the initial conditions for the universe. In this construction, the minisuperspace of 3-geometries and string vacua is a real physical configuration space for the initial conditions, rather than an abstract metauniverse of unknown structure and unknown distribution of initial patches. Decoherence among various branches/solutions on the landscape phase space, which results from the non-equilibrium quantum dynamics of matter and gravitational degrees of freedom, gives rise to a superselection rule for the universe.

\section{Why Did the Universe Start from a Low Entropy State}

As described above, we propose to take the string theory landscape to be the phase space for the initial conditions. This is justified by treating every vacuum in this field space as a potential starting point for giving birth to a Universe. An important conceptual consequence of our approach is that, as we have previously argued \cite{us1}, a picture similar to the landscape must be expected of any theory of quantum gravity. We expect quantum gravity to help with an answer on fundamental questions such as why did our universe start with these initial conditions and constants of nature. The natural follow-up question is: as compared to what other possibilities? An internal observer bound to the visible universe/box cannot meaningfully contemplate such questions. Only a so-called ``out-of-the box" observer, having access to the whole space of different possible initial conditions, quantum numbers and constants of nature can ask such questions. The 'out-of-the box' superobserver is played by the long wavelength massive modes for reasons explained in \cite{us1}. 
 
We do not have enough information about string theory at this point to do anything but construct models of the landscape (see however Ref.\cite{douglas}). In Ref.\cite{laura} one such model was provided, where the landscape was treated as a {\em disordered} lattice of vacua, where each of the $N$ sites is labelled by a mean value $\phi_i,\ i=1,\dots N$ of moduli fields. Each site has its own internal structure, consisting of closely spaced resonances around the central value. The disordering of the lattice is enforced via a stochastic distribution of mean ground state energy density $\epsilon_i,\ i=1\dots N$ of each site. These energies are drawn from the interval $\left[-W,+W\right]$, where $W\sim M_{\rm Planck}^4$ with a Gaussian distribution with width (disorder strength) $\Gamma$: $M_{\rm SUSY}^8\leq\Gamma\leq M_{\rm Planck}^8$, with $M_{\rm SUSY}$ being the SUSY breaking scale.

Quantum tunneling to other sites is always present. This allows the wavefunction to spread from site to site which, together with the stochastic distribution of sites, ensures the Anderson localization\cite{anderson} of wavepackets around some vacuum site, at least for all the energy levels up to the disorder strength. This localization forces the wavefunction to remain within the non-SUSY sector of the landscape\cite{laura}. The energy density of the Anderson localized wavepacket is $\epsilon_i =|\Lambda_i +i\gamma|$, where $\Lambda_i$ is the vacuum energy contribution to the site energy $\epsilon_i$ and $\gamma=l^{-1} l_{\rm Planck}^{-3}$, where $l$ is the mean localization length. For large enough values of the disorder strength $\Gamma$, the majority of the levels are localized so that a semiclassical treatment of their classical trajectories in configuration space is justifed. 

What does the WdW equation on this superspace look like? Initially, we restrict ourselves to FRW geometries parametrized by a scale factor $a$ and curvature parameter $k=0, 1$ for flat or closed universes, though we enlarge our superspace in the next section. The wavefunction $\Psi$ will then be a function of $a$, as well as the landscape collective coordinate $\left\{\phi_i\right\}$, which is the mean value of moduli in each lattice site. Following the usual procedure we arrive at the WdW equation\cite{wiltshire} $\hat{{\cal H}}\Psi\left(\alpha,\phi\right)=0$, with 
\be
\label{eq:wdw}
\hat{{\cal H}} = \frac{1}{2 e^{3\alpha}}\left[\frac{\partial^2}{\partial \alpha^2}-\frac{\partial^2}{\partial \phi^2} +{\cal U}(\alpha, \phi)\right],\ \ {\cal U}(\alpha, \phi)= -k e^{4\alpha}+e^{6\alpha} V(\phi).
\ee

We have followed the conventions in Ref.\cite{wiltshire}: $\alpha=\ln a$, $\phi$ is the dimensionless version of the moduli field $\Phi$, $\Phi = \sqrt{3}\slash \kappa\ \phi$ with $\kappa^2 = 4\pi G$, while ${\cal V}(\Phi) = \left(9\slash16\right) M_{\rm Planck}^4 V(\phi)$. Here ${\cal V}(\Phi)=\mu^2 \Phi^2\slash 2$ is the original modulus potential, so that $V(\phi) = m^2 \phi^2$, with $m^2 = 2\mu^2\slash \left(3\pi M_{\rm Planck}^2\right)$.
By rescaling $\phi$ as $x=e^{3 \alpha} \phi$ we can write the wavefunctional in the WdW equation:
\be
\label{eq:varsep}
\Psi(\alpha, x) = \sum_j \psi_j (x) F_j(\alpha);
\ee
inserting this decomposition into Eq.(\ref{eq:wdw}) we have:
\begin{equation}
\label{eq:sepham}
\hat{{\cal H}}_x \psi_j (x)=\epsilon_j \psi_j (x),
\end{equation}
where,
\begin{eqnarray}
&\hat{{\cal H}}_x& =\frac{\partial^2}{\partial x^2}-V(x),\nonumber \\
&\frac{\partial^2}{\partial \alpha^2}& F_j(\alpha) = \hat{\epsilon}_j \psi_j (x),
\end{eqnarray}
with $\hat{\epsilon}_j=e^{6\alpha} \epsilon_j$.

Since the disorder is large, a random matrix theory (RMT) formalism, obtained by many realizations of the stochastic potentil, was used in finding solutions to the WDW equation above. 
In Ref.\cite{us1}, we extended the landscape model of Ref.\cite{laura} to include the internal degrees of freedom in each vacuum site of our lattice. The wavefunction we use as a starting point of our analysis is found by starting from solutions of Eq.(\ref{eq:sepham}) relevant to our stochastic lattice model of the landscape and then constructing superpositions of these with a Gaussian weight that encodes the spread $v$ in energies of each lattice site due to the existence of the many moduli that act as closely spaced resonances around the mean value in each vacuum (see Ref.\cite{kiefer} for more details). If we trace out the resonances we arrive at a density matrix for this system:
\begin{equation}
\label{eq:densmat}
\rho\left(x, a; x^{\prime}, a^{\prime}\right)\sim \rho_0\left(a; a^{\prime}\right) e^{-\frac{b^2 a^2}{\pi^2}\left(x-x^{\prime}\right)^2},
\end{equation}
where $\rho_0$ is the density for the system without the resonances, $a=e^{\alpha}$ is the scale factor, and the $b=v\sqrt{M}$, where $M$ is the number of internal resonances around each vacuum. We expect $b$ to be of order the SUSY breaking scale in that vacua.

The moduli fields as well as the metric have fluctuations about their mean value and those fluctuations can serve to decohere the wavefunction\cite{hh}. This would then provide a classical probability distribution for the scaning of scale factors and moduli fields vacua. The procedure laid out in Ref.\cite{hh} starts by writing the metric and the moduli fields as
\begin{equation}
\label{eq:pert}
h_{ij} = a^2 \left(\Omega_{ij}+\epsilon_{ij}\right),\ \phi=\phi_0+\sum_n f_n(a) Q_n,
\end{equation}
where $\Omega_{ij}$ is the FRW spatial metric, $\epsilon_{ij}$ is the metric perturbation (both scalar and tensor), $Q_n$ are the scalar field harmonics in the unperturbed metric and $f_n(a)$ are the massive mode perturbations. The index $n$ is an integer for closed spatial sections, and $k=n\slash a=n e^{-\alpha}$ denotes the physical wavenumber of the mode. As stated in Ref.\cite{hh}, the fact that the CMB fluctuations are so small means that we can neglect the effects of the metric perturbations in the following calculations relative to the field fluctuations.

The wavefunction is now the functional $\Psi=\Psi(a, \phi, \left\{f_n\right\})$ on a minisuperspace which becomes infinite dimensional. Inserting Eq.(\ref{eq:pert})   into the action, yields Hamiltonians $\left\{H_n\right\}$ for the fluctuation modes, which at quadratic order in the action, are decoupled from one another. The full quantized Hamiltonian $\hat{H} = \hat{H}_0 + \sum_n {\hat{H}}_n$ then acts on the wavefunction
\be
\label{eq:wavepert}
\Psi \sim \Psi_0 (a, \phi_0) \prod_n \psi_n (a, \phi_0, f_n).
\ee 
Doing all this yields the master equation
\be
\label{eq:master}
\hat{H}_0 \Psi_0 (a, \phi_0) = \left(-\sum_n \langle  \hat{H}_n\rangle\right) \Psi_0 (a, \phi_0),
\ee
where the angular brackets denote expectation values in the wavefunction $\psi_n$ and
\be
\label{eq:nham}
\hat H_n = -\frac{\partial^2}{\partial f_n^2} + e^{6 \alpha} \left( m^2 +e^{-2 \alpha} \left(n^2-1\right)\right) f_n^2,
\ee

Following Ref.\cite{kiefer2} a time parameter $t$ can be defined for WKB wavefunctions so that the equation for the perturbations $\psi_n$ can be written as a Sch$r\ddot{\rm o}$dinger equation. If $S$ is the action for the mean values $\alpha, \phi$, define $y \equiv \left(\partial S\slash \partial \alpha\right)\slash \left(\partial S\slash \partial \phi\right)\sim \dot{\alpha}\slash \dot{\phi}$, so that we can write \cite{kiefer2}:
\bea
\label{eq:pertschr}
\psi_n &=& e^{\frac{\alpha}{2}} \exp\left(i \frac{3}{2 y} \frac{\partial S}{\partial \phi} f_n^2\right)\psi_n^{(0)}\nonumber \\
i\frac{\partial \psi_n^{(0)}}{\partial t}  &=& e^{-3 \alpha} \left\{-\frac{1}{2} \frac{\partial^2}{\partial f_n^2} + U(\alpha,\phi) f_n^2\right\} \psi_n^{(0)}\nonumber \\
U(\alpha,\phi) &=&e^{6\alpha} \left\{(\frac{n^2-1}{2})e^{-2\alpha} +\frac{m^2}{2} +\right .\nonumber \\
&+& \left . 9m^2 y^{-2}\phi^2
-6m^2 y^{-1}\phi\right\}.
\eea
During inflation, $S\approx-1\slash 3\ m e^{3\alpha} \phi_{\rm inf}$, where $\phi_{\rm inf}$ is the value of the field during inflation, so that $y=3\phi_{\rm inf}$. After inflation, when the wavepacket is in an oscillatory regime, $y$ is large so that the potential $U(\alpha, \phi)$ changes from $U_{-}(\alpha, \phi)$ to $U_{+}(\alpha, \phi)$, where 
$$
U_{\pm} (\alpha, \phi) \sim e^{6\alpha} \left[\frac{n^2-1}{2}e^{-2\alpha}\pm \frac{m^2}{2}\right].
$$
Now from Eq.(\ref{eq:pertschr}) we see that during inflation, the patches that have $U(\alpha, \phi)<0$, which can happen for small enough physical wave vector $k_n = n e^{-\alpha}$ compared to the inflaton mass term, develop tachyonic instabilities: $\psi_n \simeq e^{-\mu_n \alpha} e^{i \mu_n \phi}$, where $-\mu_n^2 = U(\alpha,\phi)f_n^2$. These trajectories {\em cannot } give rise to an inflationary universe, since they are damped in the intrinsic time $\alpha$ and so such modes do {\em not} contribute to the phase space of inflationary initial conditions. The damping of these wavefunctions in the phase space minisuperspace is correlated with the tachyonic, Jeans-like instabilities of the corresponding mode $f_n$ in real spacetime; when $U(\alpha, \phi)<0$, the solution to the equation of motion is $f_n\sim e^{\pm \mu_n t}$, while for $U(\alpha, \phi)>0$, the $f_n$ are frozen in. 

What we glean from all this is that: all initial inflationary patches, characterized by values of the scale factor $a_{\rm inf}$ and Hubble parameter $h_{\rm inf}\equiv \sqrt{2\slash 3\pi} H_{\rm inf}\slash M_{\rm Planck} $ for which $U<0$ will collapse due to the backreaction of the superhorizon modes satisfying $k_n\leq m$. Since the backreaction effects due to modes with wavenumber $n$ scale as $a^{-2}$, patches for which $U>0$ will start to inflate and the backreaction effects will be inflated away. The {\it surviving universes} are then exactly those with 
\be
\label{eq:patchinf}
m^2 \phi_{\rm inf}^2 \simeq h_{\rm inf}^2 \geq k_n^2 = \left(\frac{n}{a}\right)^2\geq m^2\Rightarrow \phi_{\rm inf}\geq 1.
\ee
We have achieved our goal, namely, the dynamics of the backreacting modes scours the Universe clean of regions which cannot support inflation! This reduction in the phase space of inflationary initial conditions implies that gravitational dynamics does {\em not} conserve the volume of the phase space, i.e. Liouville's theorem does not hold so that $\left[\hat{H}, \rho_{\rm red}\right]\neq 0$. 

We can think of the massive modes $f_n$ as collapsing into one black hole. Then we can write an  approximate expression for the entropy $S$ of the system of DeSitter patches together with the backreaction from the black hole ({\it i.e.} the massive modes), from our action including terms  up to quadratic order. This expression reduces to the entropy obtained by \cite{gh} for Schwarschild-DeSitter geometries, with the identifications
\be
S \simeq (r_I -r_{f_{n}})^2,\  r_I \simeq H_I^{-1},\  r_{f_n} \simeq H_I^{-3/2}\langle \phi_I\sqrt{U}\rangle.
\label{desitterblackhole}
\ee
where $r_I$ denotes the De-Sitter horizon of the inflationary patches with Hubble parameter $H_I$ and $r_{f_N}$ the horizon of the ``black hole'' made up from the $f_n$, where $\langle f_n\rangle \simeq \phi_{\rm inf}$, up to numerical factors of order unity.

It is interesting that the $U=0$ case, which can be thought of as a lower bound for the ``survivor'' patches, corresponds to the case of a nearly zero entropy for the de Sitter-black hole system, {\it i.e.} when the surface gravity $r_I^{-1}$ of the de Sitter patch coincides with that of the black hole, $r_{ f_{n} }^{-1}$. This means that a black hole with the same horizon as the initial inflationary patch is the borderline between the damped and survivor universes, so that the minimum entropy situation provides a lower bound on the initial conditions  $h_{\rm inf},\phi_{\rm inf}$ for an inflationary patch to appear and evolve into our universe. 

Note that in our model of the landscape as a stochastic lattice, the tracing out of the long wavelength fluctuations in the density matrix is encoded in the appearance of the localization length scale $l$ into the reduced density matrix:
\bea
\rho &=& \int \Psi(\alpha,\phi,f_n)\Psi(\alpha',\phi',f'_n) \Pi_n df_n df'_n \nonumber \\
&\simeq& \rho_0 e^{- \gamma a^6 (\phi-\phi')^2} e^{-\frac{(a\pi)^6 H^4 \mu^4(\phi-\phi')^2}{\phi_{\rm inf}^{2}} }\nonumber \\
\rho_0 &\simeq& \langle\Psi_0 (\alpha,\phi)\Psi_0(\alpha'\phi')\rangle \nonumber \\
&\simeq& e^{-M \Omega_{\rm cl} (a-a')^2}e^{-b^2 a^6 (\phi-\phi')^2}.
\label{eq:density1}
\eea
Here $\Omega_{\rm cl} = \sqrt{m_0\slash M}$, where $m_0$ sets the scale for the frequency of the internal (resonance) oscillators and $M$ is the number of internal states we traced out initially (see Ref.\cite{us1} for more details of this construction). 
\section{Holography, the arrow of time and eternal inflation}

The inflationary paradigm has been extremely succesful in explaining our observable universe. It allows us to understand both the high degree of homogeneity and isotropy of our universe while at the same time giving us a mechanism by which metric perturbations can arise from quantum fluctuations. However, at some level, it seems as if we are trading one set of fine tuning problems for the problem of what sets the initial conditions for inflation. Either the initial conditions for inflation were extremely special or else we may have to give up the possibility of ever understanding the observed arrow of time. 

We are now ready to investigate the implications of our approach to existing proposals.

\begin{itemize}

\item Eternal Inflation. 

Unfortunately, we believe that eternal or chaotic inflation models reintroduce the problem of the initial conditions as forcefully as ever, albeit in a disguised form. How do we calculate the probability of each bubble? Although every bubble that arises via a quantum fluctuation at any energy scale has a non zero probability to form, calculating the probability of each pocket universe according to the equilibrium formula, $P=\exp S_E$ simply states that high energy pockets remain the most unlikely ones \cite{albrecht} while the low energy pockets run into trouble with the arrow of time. This brings us back to the basic problem and making more bubbles of these bubbles does not  solve the problem.  We believe that the reason lies in the hidden assumption of ergodicity and equilibrium made for the state of each bubble; making these assumptions leads us to the troublesome expression $P=\exp S_E$. As we have shown, ergodicity and equilibrium are broken in each pocket  due to the quantum dynamics of gravity.\footnote{An interesting proposal for the probability distribution in eternal inflation scenarios can be found in \cite{easther}} 

\item Holography-inspired proposals.

The success of holography in explaining the statistical mechanics behind the area law for the entropy of a black hole has inspired an investigation of its impact on cosmology and especially on the selection of the initial conditions \cite{bousso}.In analogy with black holes a principle of complementarity has been put forth for cosmology \cite{causalpatch}. Some of the proposals put forth, such as casual patch physics\cite{causalpatch}, N and D-bound \cite{bousso}, etc. require that the degress of freedom of the cosmological system be confined to the area $A$ rather than the volume. This requirement therefore excludes the possiblity of any entaglement between the in and out modes of our visible universe. The various proposals differ from each other on whether to include the area $A$ of the diamond shaped past and /or future casual horizons. However, they all agree on the statement that entropy $S/A <1$. These proposals seem to become even more problematic at late-times for $\Lambda CDM$ cosmologies, since $\Lambda$ introduces a cutoff of the area and thus an upper bound on the entropy of the casual patch via the relation $A \simeq 3\pi / \Lambda$. We would like to point out that confining the information to the surface area $A$ rather than the enclosed volume $V$ of the visible universe does not solve the problem of the selection of the initial conditions for the following reason: the probability of that initial patch is still estimated by assuming the equilibrium expression for the horizon degrees of freedom $P=\exp S_E$, on the basis of phase space ergodicity and no entanglement for the causal patch. As shown in the previous section and in \cite{us1}, this is not the case! The initial mixed state can not evolve to a pure state under a unitary evolution; the system is not ergodic because, as we showed, the reduced density matrix evolves with time, $[H,\rho_{red}]=d\rho/dt$. Giving up the assumptions of ergodicity,equilibrium and lack of entanglement for the system thus places severe doubts, at a fundamental level, on the validity of holographical approaches such as causal patch physics, complementarity principle and cosmological holography.We can see from Eqn.(\ref{eq:density1}) for example that some entanglement with the super-Hubble length modes survives. Our universe remains in a mixed state. Physically, super-Hubble modes re-enter the horizon in an excited state\cite{gh} and thus carry with them the memory of the initial entanglement on the inflaton and backreaction effects.  Even if the degree of entanglement is small at present, it will never allow the system to evolve to a pure state in thermal equlibrium thereby have all of its degrees of freedom contained on the surface area of the casual patch. As it is clear from the time evolution of the density matrix, the nonergodic phase space does not preserve its volume, thereby neither does the the Hilbert space available to our causal patch. What about arguments based on Poincarre recurrence that lead to paradoxes when applied to a late time $\Lambda$ dominated universe? Again, a recurrence fluctuation that would take us close enough to the initial state, can be applied only when the phase space is conserved. If the volume of phase space is dynamically compressed then, no matter how long we wait, we may never return close enough to the starting point in the phase space. In short,no fluctuation may take us through a Poincarre cycle, if the phase spaces is not ergodic. Simply put, unlike stationary configurations such as black holes, expanding cosmologies can never reach equilibrium. Subsequently there is a flow of information accross the horizon as a result of the universe wavepacket remaining in a mixed state. Making holography compatible with cosmology has proven to be a notoriously challenging problem. Our analysis here indicates that the difficulty may not be technical but rather conceptual and hidden in the underlying assumptions of the theory. We hope that our investigation of the quantum dynamics of gravitational degrees of freedom has shed light into the reasons why holography and causal patch physics may not be applicable to cosmology. All trouble stems from the fact that the two pillar assumptions for the applicability of these approaches, namely, equilibrium and ergodicity, are not warranted for systems that contain both matter and gravitational degrees of freedom, as exhibited in the previous section.

\item  What is the highest energy scale of Inflation?

What happens if the initial fluctuation from the minimum $\phi_{\rm inf}$ is much larger than its lower bound? Is there an upper bound on the scale of inflation? This is a difficult question to address since $\phi_{\rm inf} \gg l$ marks the breakdown of our semiclassical treatment and takes us deep into the quantum gravity regime. Nonetheless, we can extract some information by trying to extend our analysis to these cases. We have argued that the higher the scale of inflation, the less significant the backreaction of massive perturbations becomes. It follows that in the regime of extremely high energy inflation, the backreaction term becomes negligble, equivalently expressed by $[\hat{H},\rho]\approx 0$. Assuming that quantum mechanics remains a valid theory in the transplanckian regime, then arguments based on Poincare recurrence phenomena may now be valid in this regime only.  But in this case, the Poincare recurrence time, $T_{\rm recurr}\simeq e^{S}$ is short so that these patches become quantum on times scales of order $T_{\rm recurr}$ by a broadening of their energy level given by $\Delta E \simeq T^{-1}_{\rm recurr}$. The upper bound on the energy scale of inflation is given by the requirement that the recurrence time of the patch should allow for enough efoldings and be much larger than the age of the universe. For short recurrence time, the broadening of the energy level of the inflating patch becomes larger than the spacing between levels which in turn, results in the loss of classicality and the universe becoming quantum rather then inflating. The requirement on recurrence time thus translates into a constraint on the energy scales of inflation, namely the quantum entanglement must occur over long enough times such that it allows inflation to take its course. In Section 2 we showed how the lower bound on the energy scale of inflation is obtained by the requirement that the universe survives the strong influence of massive modes backreaction on the geometry and continues to expand, a scenario corresponding to the minimum possible entropy. In this part we derived an upper bound on the initial patch energy,by the constraint that the univese should not become quantum before inflation ends. This  upper bound is obtained by showing that, patches with energy scales larger than the fundamental scale,(here given by $M_p^4 \simeq \gamma$), have a recurrence time of order one in fundamental units in which case the universe loses its classicality by going through a Poincarre cycle before inflation ends. We conclude that the bounds on the initial inflationary patches give a range of $O(M_{susy})\le \phi_i\le O(\sqrt{\gamma})$. In this range the superselection rule emerges from the out-of-equilibrium quantum dynamics of the gravitational and matter degrees of freedom described in Section 2.
\end{itemize}

\section{Discussion}
Why did the Universe start in a state of lower than expected entropy? Equivalently, how did high scale inflation occur? The key to answering these questions is to not be fooled by arguments based on {\em equilibrium} statistical mechanics. In fact, it is exactly the {\em non}-equilibrium dynamics of superhorizon modes and their backreaction onto the mean values of  $a,\ \phi$ that selects out the regions which inflate; patches that do not satisfy $m<H_I, l<\phi_I<b$ will recollapse. This non-equilibrium dynamics also leads to non-ergodic behavior in the phase space of initial conditions, as well as entanglement of states. A unitary evolution will not allow the mixed state of the initial state evolve later to a pure state thus some degree of entanglement is always present. Our proposal is based in studying the quantum dynamics of gravity by non-equilibrium methods.  

What are the implications of our findings to the current proposals in literature for the selection of the initial conditions and the emergence of an arrow of time?  Unlike the black hole geometries, applying holography to cosmological scenarios has proven very hard. Proposals of causal patch physics and eternal inflation for the selection of the initial conditions seem to run into paradoxes since they can not simultaneoulsy accomodate a high scale inflation with the emergence of a thermodynamic arrow of time. Neither can they resolve the entropy bound that a low scale cosmological constant introduces on the energies of the initial conditions. The paradoxes stemming from this proposals make the relation between gravity,thermodynamics and quantum mechanics obscured since,within the framework of the proposals, the underlying principles of causality, the second law of thermodynamics and unitarity appear at odds with each other.
As is common in physics, often paradoxes indicate that some of the assumptions we take for granted about the basic principles are not justified. Here we demonstrated that two of the hidden assumption on which these proposals are based, namely an equilibrium dynamics for the initial patches and ergodicity of the phase space, are not valid. 

Under a unitary evolution, our findings of a non-equilibrium dynamics and nonergodic phase space are significant, since they imply that a holographic description of gravity during inflation may not be tenable. By removing the assumption of equilibrium and ergodicity, our analysis here sheds light into the roots of some of the paradoxes. But it also gives rise to questions about the applicability of the principle of complementarity, causal patch physics and, the capability of the eternal inflation scenario to provide a satisfactory and coherent answer with respect to the selection of initial conditions and the understanding of the arrow of time. 

Despite having made use of a particular model of the landscape to arrive at our results we would argue that our results should have wider applicability. The landscape minisuperspace serves mostly to provide a concrete realization of our approach, specifically the scales $M_*, M_{\rm SUSY}$ for the widths of the initial inflationary patches. The rest of the quantum cosmological calculation based on backreaction and the master equation is general and could be applied to any phase space for the initial conditions once its structure was known.

Can we test these ideas? In a forthcoming paper\cite{richlauranext} we will report how remnants of quantum entaglement between in and out modes, as represented by the cross-terms in the reduced density matrix, might be tested by cosmological observables such as nongaussianities in CMB and large scale structure.

\ack
R.H. was supported in part by DOE grant DE-FG03-91-ER40682. LMH was suported in part by DOE grant DE-FG02-06ER41418 and NSF grant PHY-0553312.

\end{document}